\documentclass[structabstract]{aa}
\usepackage{graphicx}
\usepackage{natbib}\bibpunct{(}{)}{;}{a}{}{,} 
\usepackage{txfonts}
\begin{document}
   \title{Core properties of $\alpha$ Cen A using asteroseismology}


   \author{P. de Meulenaer\inst{1},
          F. Carrier \inst{2}, 
          A. Miglio\inst{1},
          T.R. Bedding\inst{3},
	  T.L. Campante\inst{4,5},
          P. Eggenberger \inst{6,1},
          H. Kjeldsen \inst{4},
          and J. Montalb\'an \inst{1}
          }

   \institute{Institut d'Astrophysique et de G\'eophysique de l'Universit\'e de Li\`ege, All\'ee du 6 Ao\^ut, 17 B-4000 Li\`ege, Belgium\\
             \email{[miglio;eggenberger;montalban]@astro.ulg.ac.be}
         \and
             Instituut voor Sterrenkunde, Katholieke Universiteit Leuven, Celestijnenlaan 200D, 3001 Leuven, Belgium\\
             \email{fabien@ster.kuleuven.be}
         \and
             Sydney Institute for Astronomy (SIfA), School of Physics A28, University of Sydney, NSW 2006, Australia\\
             \email{Bedding@physics.usyd.edu.au}
         \and
             Danish AsteroSeismology Centre (DASC), Department of Physics and Astronomy, Aarhus University, DK-8000 Aarhus C, 		Denmark\\
             \email{[hans;campante]@phys.au.dk}
         \and
	     Centro de Astrof{\'\i}sica da Universidade do Porto, Rua das Estrelas, 4150-762 Porto, Portugal
         \and
             Observatoire de Gen\`eve, Universit\'e de Gen\`eve, 51 chemin des Maillettes, CH-1290 Sauverny, Switzerland
             }



\abstract
{A set of long and nearly continuous observations of $\alpha$ Centauri A should
allow us to derive an accurate set of asteroseismic constraints to compare to
models, and make inferences on the internal structure of our closest
stellar neighbour.}
{We intend to improve the knowledge of the interior of $\alpha$ Centauri A
by determining the nature of its core.}
{We combined the radial velocity time series obtained in May 2001 with
  three spectrographs  in Chile
and Australia: CORALIE, UVES, and UCLES. 
The resulting combined time series has a length of 12.45 days
and contains over 10,000 data points and allows to greatly reduce
the daily alias peaks in the power spectral window. }
{We detected 44 frequencies that are in good overall agreement with previous
studies, and found that 14 of these show possible rotational splittings. New
values for the large ($\Delta\nu$) and small separations ($\delta\nu_{02},
\delta\nu_{13}$) have been derived.}
{A comparison with stellar models indicates that the asteroseismic constraints
determined in this study (namely $r_{10}$ and $\delta\nu_{13}$) allows us
to set an upper limit to the amount of convective-core overshooting needed
to model stars of mass and metallicity similar to those of $\alpha$ Cen A.}

   \authorrunning{P. de Meulenaer et al.}

   \maketitle
%

\section{Introduction}
During the last decade, the visual binary stellar system $\alpha$ Centauri
turned out to be a very interesting asteroseismic target to observe and
model because of its proximity and of the similarity of these stars to the
Sun. Moreover, the mass of the primary component is very close to the limit
above which stellar models predict the onset of convection in the
energy-generating core.  This makes the theoretical study of $\alpha$ Cen A
particularly valuable for testing the poorly-modelled treatment of convection
and extra-mixing in the central regions of low-mass stars.

The proximity of $\alpha$ Cen ($d$\,=\,1.34\,pc) provides a relatively well
determined parallax, $\pi$\,=\,747.1\,$\pm$\,1.2\,mas
\citep{Soderhjelm99}. This visual binary system shows an eccentric orbit
($e = 0.519$) with a period of almost 80 years and so the masses of the
two components are very well constrained. The more luminous component, 
$\alpha$ Cen A, is a G2V star with a mass of $1.105 \pm 
0.007$\,M$_{\odot}$.   The secondary, $\alpha$ Cen B, is a cooler K1V
star with a mass of $0.934 \pm 0.006$\,M$_{\odot}$
\citep{Pourbaix02}, so they bracket the Sun in mass. 
Their effective temperatures are
$T_{\rm eff,A} = 5810 \pm 50$\,K and $T_{\rm eff,B} = 5260
 \pm 50$ K \citep[see][]{Porto08}. \citet{Kervella03} have measured
the angular diameters of $\alpha$ Cen A and B with
VINCI/VLTI. Combining with the parallax gave linear radii for
the two stars: $R_{A} = 1.224 \pm 0.003 {\rm R}_{\odot}$ and
$R_{B} = 0.863 \pm 0.005 {\rm R}_{\odot}$. Masses and radii allow
determination of the surface gravities: $\log g_{A} = 4.307 \pm 
0.005$ and $\log g_{B} = 4.538 \pm 0.008$, an accuracy rarely
reached in stars other than the Sun. Moreover, thees brightness of both
components of the system ($V_{A} = -0.01$ and $V_{B} = 1.33$) allows the
acquisition of extremely high-quality spectra.





Asteroseismic data have been obtained by several teams. The first unambiguous
detection of p-modes in $\alpha$ Cen A was made by \cite{Bouchy01,Bouchy02}
during a 13-night campaign with the spectrograph CORALIE, confirming the
earlier claimed detection made by \cite{Schou00} with the WIRE
satellite. \cite{Bouchy02} detected 28 modes with angular degrees
$\ell = 0,1$ and $2$. At the same time, another team
\citep{Butler04,Bedding04} made observations during five nights from Chile
with UVES and from Australia with UCLES, and detected $42$ frequencies with
$\ell = 0,1,2$ and $3$. \cite{Kjeldsen05} also provided a value of the mode
lifetime for $\alpha$ Cen A of $2.3_{-0.4}^{+1.0}$ days. \cite{Fletcher06}
then carried out a re$-$analysis of the WIRE observations, resulting in
additional asteroseismic data. In particular, they suggested two values of the
rotational frequency, $0.54 \pm 0.22 {\rm \mu Hz}$ and $0.64 \pm
 0.25 \: {\rm \mu Hz}$, using two different analysis methods, 
and a mode lifetime of $3.9 
\pm 1.4$ days. More recently, \citet{Bazot07} detected $34$ modes with
the HARPS spectrograph in Chile, and suggested five rotational splittings
for $\ell = 2$ modes. Asteroseismic data have also been obtained for the B
component \citep{Carrier03,Kjeldsen05}.



The $\alpha$ Cen system has been also extensively
modelled. \citet{Guenther00}, \cite{Morel00} and \cite{Noels91} performed a
calibration based only on non$-$asteroseismic constraints. They found that two
kinds of models, one with a radiative core and the other with a convective
one, could satisfy these constraints. Subsequently, the high-quality
non$-$asteroseismic constraints together with asteroseismic constraints stimulated new
calibrations of the stellar system \citep{Thevenin02,
Thoul03,Eggenberger04,Miglio05}. Although these authors succeeded to fit
some of the asteroseismic constraints, they were not in a position to draw a
definite conclusion on the convective versus radiative nature of the core
of $\alpha$ Centauri A. To perform such a discrimination, all teams
suggested that more accurate asteroseismic constraints were needed.

In the present paper, the May 2001 data of \cite{Bouchy02} from CORALIE and of
\cite{Bedding04} from UVES and UCLES have been unified to compute a
combined velocity time series in order to reduce the effect of daily
aliases in the power spectrum. The objective is to provide a more accurate
set of asteroseismic constraints that allows a better comparison with theoretical
models and a clear discrimination between them.

Section 2 describes the different data sets and their main features, along
with the method used to analyse the acoustic spectrum. Section 3 presents
the frequencies and amplitudes detected, along with new values for the large
and small separations and a rotational frequency. In the section 4, we
compare these new observational constraints with the models of $\alpha$ Cen
A presented by \cite{Miglio05}. Section 5 is dedicated to the conclusions.


\section{Data and methods}

   \begin{figure} 
   \resizebox{\hsize}{!}{\includegraphics{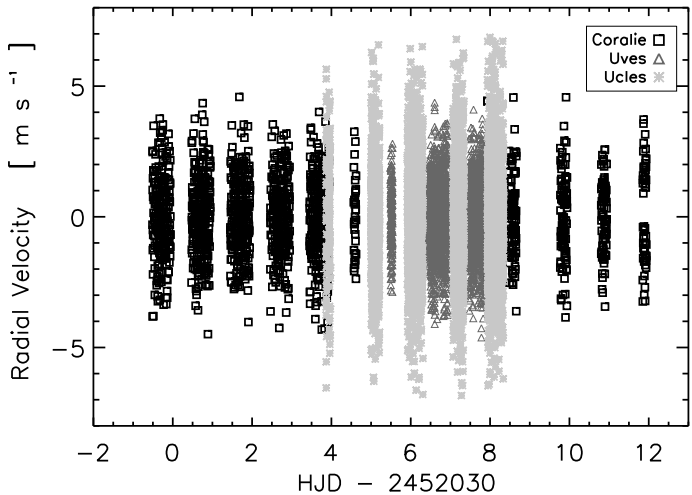}}
   \caption{Combined time series of CORALIE, UVES and UCLES. One can see
   that the UCLES data, which were taken in Australia, fill several gaps in
   the time series of the CORALIE and UVES data, taken in Chile. This will
   allow a better detection of p-modes frequencies by reducing the daily
   aliases in the spectrum of the star (see text for details).}
   \label{fig:ts}
   \end{figure}

\subsection{Data sets}
$\alpha$ Cen A was observed in May 2001 in Chile by \citet{Bouchy02} during
a 13-night campaign with the CORALIE fiber$-$fed \'echelle spectrograph,
mounted on the 1.2 m Euler Swiss telescope at the ESO La Silla Observatory
\citep{Bouchy01,Bouchy02}. Another team \citep{Butler04,Bedding04} made
two-site observations over five nights from Chile and 
Australia. They used UVES (UV$-$Visual Echelle Spectrograph) at the 8.2 m
Unit Telescope 2 (Kueyen) of the VLT (Very Large Telescope) in Chile, and
UCLES (University College London Echelle Spectrograph) at the 3.9 m AAT
(Anglo-Australian Telescope) at Siding Spring Observatory in
Australia. These two spectrographs were provided a stable wavelength
reference using an iodine cell \citep[see][]{Butler96}. 

The median cadence of the data set was one spectrum every 150\,s for  CORALIE,
26\,s for UVES and 20\,s for UCLES\/.  The durations of
the observations and the total number of spectra are given in
Table~\ref{tab:features}.

Figure \ref{fig:ts} shows the times series of the different
spectrographs. One can see in Tables \ref{tab:features} and \ref{tab:noise}
that each time series has its own advantages.  CORALIE has the longest time
series, and thus the best frequency resolution ($0.93 {\rm \mu Hz}$). UVES
data have the best signal-to-noise ratio in the power spectrum (see
Table~\ref{tab:noise}). One can also note (Fig.~\ref{fig:ts}) that the
UCLES nights, observed from Australia, fill some gaps in the CORALIE and
UVES nights, observed from Chile. This is important when 
combining the data, as discussed below.


\begin{table*}
\centering
\caption{\small Table of the main features of the three campaigns. $^{1}$1.2m ESO, $^{2}$8.2m VLT, $^{3}$3.9m AAT}
\begin{tabular}{ccccccccccccc}
\hline
\hline
Campaign & Duration of observation [days] & Frequency resolution [${\rm \mu Hz}$] & median cadence [$s$] & number of spectra \\
\hline
CORALIE$^{1}$ & $12.45$ & 0.93 & $150$ & $1850$\\
UVES$^{2}$    & $2.42$  & 4.78 & $26$  & $3013$\\
UCLES$^{3}$   & $4.47$  & 2.59 & $20$  & $5169$\\
\hline
 \end{tabular}
\label{tab:features}
\end{table*}			

   \begin{figure}
   \resizebox{\hsize}{!}{\includegraphics{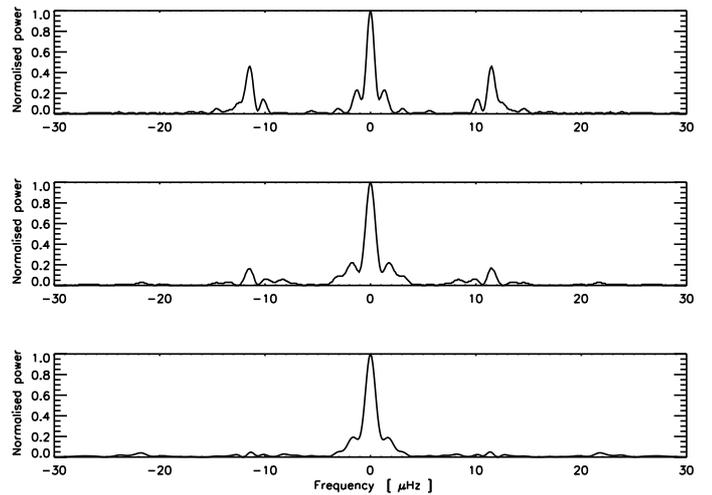}}
   \caption{Comparison of the spectral windows of the time series of the
   CORALIE data (top panel) and the one of the combined time series of the
   CORALIE, UVES and UCLES data with standard weights (middle panel) and
   sidelobe-optimised weights (bottom panel). The daily aliases
   are already much reduced in the case of the combined time series with
   standard weights, as expected from the fact that UCLES nights fill
   several gaps in the CORALIE and UVES time series.}
   \label{fig:spw}
   \end{figure}

\subsection{Combining data sets and weighting}

Combing the different velocity time series allows us to reduce the aliases
at 1\,d$^{-1}$ (or 11.57 ${\rm \mu Hz}$) that arise from the daily gaps in
the time series.  The top panel of Fig. \ref{fig:spw} shows the spectral
window of the CORALIE data alone and the middle panel shows the result for
the combined time series, using standard weights (see below).  The
1\,d$^{-1}$ alias peaks have been reduced by a factor $\sim 2.6$ in power
between the two.  



\begin{table*}
\centering
\caption{\small Table of the level noise (in the power spectrum) for the
different campaigns and their combination $^{1}$[\citet{Bouchy02}],
$^{2}$[\citet{Butler04}], $^{3}$[\citet{Bedding04}], $^{4}$ This
paper. Standard, sidelobe-optimised and noise-reduced stand for the different
weights used in the power spectra.}
\begin{tabular}{ccccccccccccc}
\hline
\hline
Campaign & $\sigma_{pow}$ [m$^2$\,s$^{-2}$] & $\sigma_{amp}$ [cm\,s$^{-1}$] & $3\times$ S/N [cm\,s$^{-1}$]\\
\hline
CORALIE (standard)$^{1}$ & $2.39 \times 10^{-3}$ & $4.3$ & $12.9$\\
UVES (standard)$^{2}$ & $6.91 \times 10^{-4}$ & $2.33$ & $7.0$ \\
UVES (noise-optimised)$^{2}$ & $5.67 \times 10^{-4}$ & $2.11$ & $6.3$ \\
UCLES (standard)$^{2}$ & $3.17 \times 10^{-3}$ & $4.99$ & $15.0$ \\
UCLES (noise-optimised)$^{2}$ & $2.43 \times 10^{-3}$ & $4.37$ & $13.1$ \\
UVES-UCLES (sidelobe-optimised)$^{3}$ & $1.07 \times 10^{-3}$ & $2.8$ & $8.7$ \\
CORALIE-UVES-UCLES (standard)$^{4}$ & $8.58 \times 10^{-4}$ & $2.59$ & $7.8$  \\
CORALIE-UVES-UCLES (sidelobe-optimised)$^{4}$ & $1.03 \times 10^{-3}$ & $2.84$ & $8.5$  \\
CORALIE-UVES-UCLES (noise-optimised)$^{4}$ & $6.56 \times 10^{-4}$ & $2.27$ & $6.8$  \\
\hline
\end{tabular}
\label{tab:noise}
\end{table*}		

\begin{figure}
   \resizebox{\hsize}{!}{\includegraphics{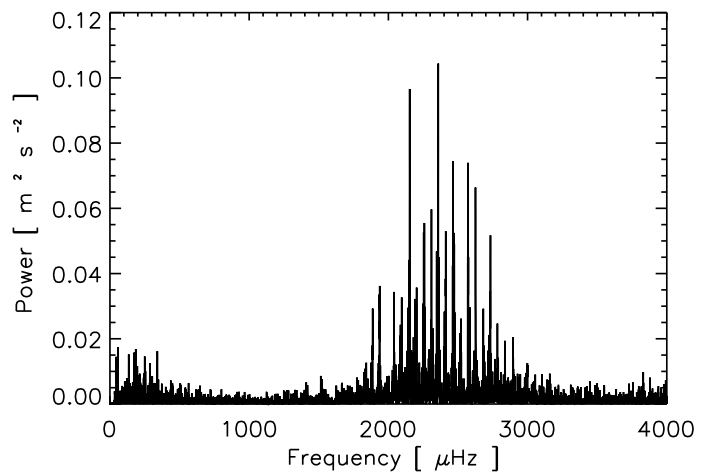}}
   \caption{Power spectrum of the combined time series of CORALIE, UVES and
   UCLES (standard weights).}
   \label{fig:pws}
   \end{figure}

We used the Lomb-Scargle modified algorithm, suited for unevenly spaced
data (\cite{Lomb76}, \cite{Scargle82}), to compute the power spectrum of
the combined velocity time series.
We used three different weighting schemes, which we will refer to as
standard weights, sidelobe-optimised weights and noise-optimised weights:
\begin{enumerate}

\item We used the measurement uncertainties,
$\sigma_{i}$, as weights in calculating the power spectrum, which is
displayed in Fig. \ref{fig:pws}. With these so-called standard weights $w_{i} =
1/\sigma_{i}^{2}$, the mean white noise level in the power spectrum,
computed between 7.5 and 15 mHz, is $8.58 \times
10^{-4}$\,m$^{2}$\,s$^{-2}$.  Assuming the noise to be white
($\sigma_{amp} = \sqrt{\frac{\pi}{4} \sigma_{pow}}$), the mean noise level
in the amplitude spectrum is $2.59$\,cm\,s$^{-1}$.

\item We applied the method described by \citet{Bedding04} in which weights
are optimised in order to reduce the daily aliases.  We adjusted the
weights on a night-by-night basis in order to optimise the window
function. To be specific, we allocated separate adjustment factors to each
of the CORALIE, UVES and UCLES nights. The weights on each night were
multiplied by these factors and the spectral window was calculated, and
this process was iterated to minimize the height of the sidelobes.  The
spectral window of the sidelobe-optimised spectrum is displayed in
Fig. \ref{fig:spw} (bottom panel).

\item The third power spectrum was computed by means of noise-optimised
weights, as described by \citet{Arentoft08}.  The short-term variations in
the uncertainty time series were first removed by bandpass-filtering. This
removed fluctuations in the weights on the timescale of the stellar
oscillations.  We also studied the velocity residuals in which both the
slow variations and the stellar oscillations were removed.  We compared the
velocity residuals, $r_i$ with the corresponding uncertainty estimates,
$\sigma_i$. Bad data points are those for which the ratio $|r_i/\sigma_i|$
is large, i.e., where the residual velocity deviates from zero by more than
expected from the uncertainty estimate.  Bad data points were then
down-weighted by dividing all $\sigma_i$ by $\sqrt{f(x_i)}$, where $f(x_i)$
is an analytical function:
\begin{equation}
  f(x_i) = \frac{1}{1+(x_i / x_0)^{10}} , x_i = |r_i / \sigma_i| . 
\end{equation}
The adjustable parameter $x_0$ controls the amount of down weighting; it
sets the values of $|r_i / \sigma_i|$ for which the weights are multiplied
by 0.5, and so it determines how bad a data point should be before it is
down weighted. The optimum choice for $x_0$ was found through
iteration. For each trial value of $x_0$, the noise level was determined
from a time series in which all power had been removed at both the
low-frequency noise and the high-frequency oscillations. We chose the $x_0$
that resulted in the lowest noise in the power spectrum.  These weights
allowed us to diminish the level of noise (See Table~\ref{tab:noise}).

\end{enumerate}

\subsection{Frequency analysis}
\label{p-modes extraction}

The method used to extract the modes was the standard pre-whitening
procedure. It consists of locating the highest peak in the region that
contains the oscillation frequencies and subtracting the corresponding
sinusoid from the time series. We then computed a new power spectrum,
located the highest peak and iterated the procedure until no peaks remained
above S/N = 3.


When a peak in the power spectrum is removed by subtracting the
corresponding sinusoid from the time series, the daily aliases associated
with this peak also disappear from the power spectrum. This works well for
high-amplitude peaks but at low amplitudes the reinforcement by noise may
cause an alias peak to be mistaken for the real peak.  By analysing three
versions of the power spectrum, we hoped to reduce this problem.  It should
also be kept in mind that if the oscillation modes are resolved, i.e., the
oscillations have a shorter lifetime than the duration of the observations,
then each oscillation mode will produce multiple peaks.

The uncertainties on the frequencies were computed using the formula of
\citet{Libbrecht92}:
\begin{equation}
\sigma_{nl}^{2} = \frac{\Gamma}{4\pi T}\sqrt{\beta + 1}(\sqrt{\beta + 1} + \sqrt{\beta})^{3}  \ \ ,
\label{eq:Libbrecht92}
\end{equation} 
where $T$ is the total length of the time series, $\beta$ is the reciprocal
of the signal-to-noise ratio of the mode, and $\Gamma$ is the line-width of
the mode.  Note that $\Gamma$ is related to the mode lifetime ($\tau$) via
$\Gamma = 1/\pi \tau$.  In fact, the modes are not resolved in the spectrum
and so we set $\Gamma$ to the frequency resolution, $0.93 \: {\rm \mu Hz}$,
which is the same value that would be deduced from the lifetime of $3.9 \pm
1.4$\,d reported by \cite{Fletcher06}.

   \begin{figure}
   \resizebox{\hsize}{!}{\includegraphics{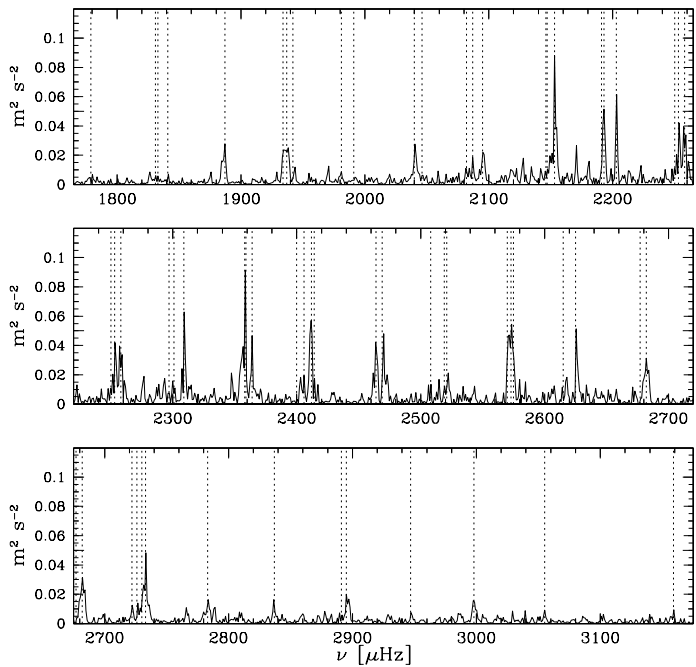}}
   \caption{Power spectrum of $\alpha$~Cen~A. The frequencies in
   Table~\ref{tab:noise} are indicated with dotted lines.}
   \label{fig:zoom}
   \end{figure}

\section{Results}

\subsection{Mode identification and \'echelle diagram}
\label{mode_id_ech_diag}

In solar-like stars, p-mode oscillations produce a characteristic comb-like
structure in the power spectrum, which is well-approximated by the
asymptotic relation \citep{Tassoul80}:
\begin{equation}
\nu_{n,\ell} \approx \Delta \nu_{0}(n+\frac{\ell}{2}+ \varepsilon ) -
\ell(\ell+1) D_0, 
\label{eq:asympt_relation}
\end{equation} 
where $\Delta \nu_{0} = \left\langle \nu_{n,\ell} - \nu_{n-1,\ell}
\right\rangle $ and $D_0$ $\approx$ $\frac{1}{6}\delta \nu_{02} =
\frac{1}{6} \left\langle \nu_{n,0} - \nu_{n-1,2} \right\rangle $.  The two
quantum numbers $n$ and $\ell$ correspond to the radial order and angular
degree, respectively. Since the stellar disk is not resolved, only the
lowest degree modes ($\ell \leq 3$) can be detected.
In case of stellar rotation, a third quantum number,~$m$, describes
the splitting of the frequencies:
\begin{equation}
\nu_{n,\ell,m} \approx \nu_{n,\ell,0} + m \Omega,
\label{eq:rotation}
\end{equation}
with $-\ell \leq m \leq \ell$ and with $\Omega$ being the frequency of the
rotation.

\begin{figure}
   \resizebox{\hsize}{!}{\includegraphics{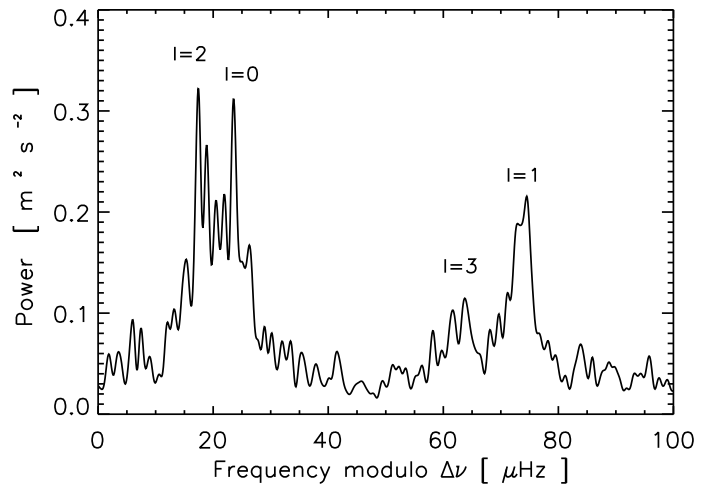}}
   \caption{Collapsed \'echelle diagram. The first major peak, below $20 \:
   {\rm \mu Hz}$ is an $\ell = 2$ and the second, at $25 \: {\rm \mu Hz}$
   is an $\ell = 0$. The two peaks around $65 \: {\rm \mu Hz}$ belongs to
   the $\ell = 3$ mode and the last peak, at $75 \: {\rm \mu Hz}$ is an
   $\ell = 1$.}
   \label{fig:sed}
   \end{figure}

In the power spectrum of $\alpha$ Cen A, displayed in
Fig. \ref{fig:pws} (computed with the standard weights), we see a
series of peaks between 1.8 mHz and 3 mHz, as expected for
solar-like oscillations in a star of this mass and radius.  A
periodicity of about $106 \: {\rm \mu Hz}$ is clear in the power
spectrum, a fact that can also be seen in its autocorrelation. In order to
identify the angular degree $\ell$ of each mode, the power spectrum has
been cut in slices of $106 \: {\rm \mu Hz}$ and summed, in order to
build the collapsed \'echelle diagram (Figure \ref{fig:sed}). This diagram
shows peaks corresponding to $\ell \: =$ 0,1,2 and 3 modes. We
also see in this figure that the side-lobes due to aliases, which often
complicate the analysis, are significantly reduced. This collapsed
\'echelle diagram allows us to identify clearly the modes $\ell$ = 0,1,2
and the possible presence of $\ell$ = 3 modes, and to reject other
frequencies that were obviously false p-modes extractions. Our
identification agrees with those previously published.

   \begin{figure}
   \resizebox{\hsize}{!}{\includegraphics{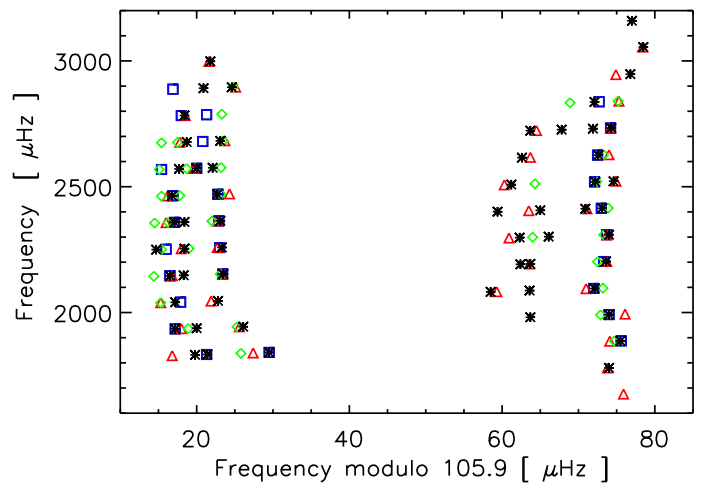}}
   \caption{Echelle diagram of the new frequencies (black asterisks) and
   those reported by earlier studies: CORALIE \citep{Bouchy02} (blue
   squares), UVES-UCLES \citep{Bedding04} (red triangles), HARPS
   \citep{Bazot07} (green diamonds).}
   \label{fig:echelle_diag}
   \end{figure}




\begin{table*}
\centering
\caption{\small Table of the frequencies detected (in ${\rm \mu Hz}$). The
  error bars have been computed by the formula \ref{eq:Libbrecht92}.}   
\begin{tabular}{ccccccccccccc}
\hline
\hline
 & $\ell = 0$ & $\ell = 1$ & $\ell = 2$ & $\ell = 3$\\
\hline
n = 15 &	      		  &	$1779.9 \pm 0.6$ 		  		  &	$1831.6 \pm 0.6$/$1833.2 \pm 0.5$		&					 \\
n = 16 & $1841.3 \pm 0.6$ &	$1887.3 \pm 0.5$ 		  		  &	$1934.9 \pm 0.5$/$1937.7 \pm 0.5$		& $1981.4 \pm 0.7$	\\
n = 17 & $1942.4 \pm 0.6$ &	$1991.7 \pm 0.6$ 		  		  &	$2040.8 \pm 0.5$						& $2082.1 \pm 0.5$/$2087.2 \pm 0.6$	\\
n = 18 & $2046.4 \pm 0.6$ &	$2095.7 \pm 0.5$ 		  		  &	$2146.0 \pm 0.5$/$2147.8 \pm 0.5$		& $2191.9 \pm 0.5$/$2193.2 \pm 0.5$	\\
n = 19 & $2153.0 \pm 0.4$ &	$2203.1 \pm 0.5$ 		  		  &	$2250.1 \pm 0.5$/$2253.8 \pm 0.5$		& $2297.7 \pm 0.5$/$2301.5 \pm 0.5$	\\
n = 20 & $2258.7 \pm 0.4$ &	$2309.4 \pm 0.4$ 		  		  &	$2358.3 \pm 0.4$/$2359.6 \pm 0.5$		& $2400.7 \pm 0.7$/$2406.3 \pm 0.6$	\\
n = 21 & $2364.4 \pm 0.5$ &	$2412.1 \pm 0.5$/$2414.4 \pm 0.6$ &	$2464.0 \pm 0.5$						& $2508.4 \pm 0.5$	\\
n = 22 & $2469.9 \pm 0.5$ &	$2519.4 \pm 0.5$/$2521.8 \pm 0.5$ & $2570.8 \pm 0.5$/$2573.1 \pm 0.5$		& $2615.7 \pm 0.5$	\\
n = 23 & $2575.2 \pm 0.6$ &	$2625.7 \pm 0.4$		  		  & $2677.7 \pm 0.5$						& $2722.7 \pm 0.6$/$2726.8 \pm 0.6$	\\
n = 24 & $2682.1 \pm 0.5$ &	$2730.9 \pm 0.6$/$2733.3 \pm 0.4$ &	$2783.3 \pm 0.5$						&					 \\
n = 25 &          	      &	$2837.0 \pm 0.6$		  		  & $2891.7 \pm 0.6$						&					 \\
n = 26 & $2895.4 \pm 0.5$ &	$2947.6 \pm 0.6$		  		  & $2998.5 \pm 0.6$						&					 \\
n = 27 &	      		  &	$3055.2 \pm 0.6$		  		  &  										&					 \\
n = 28 &	      		  &	$3159.6 \pm 0.6$		  		  &											&					 \\

\hline
\end{tabular}
\label{tab:freq}
\end{table*}

Table \ref{tab:freq} and Fig.~\ref{fig:zoom} give the 44 modes that
have been detected. For the particular case of the $\ell$ = 3 modes, we
paid attention to the proximity of these modes to the daily alias peaks of
the $\ell$ = 1 modes. 
The sidelobe-optimised power spectrum, where the aliases
were further diminished by optimising the weights, helped us to ensure that
the peaks detected were $\ell$ = 3 modes and not aliases of $\ell$ = 1
modes.

The \'echelle diagram displayed in Fig.~\ref{fig:echelle_diag} is a
convenient tool to illustrate the properties of a solar$-$like star. It was
introduced by \cite{Grec83} to identify the degree $\ell$ of solar
modes. It is computed by plotting the detected frequencies 
modulo the large separation $\Delta\nu$. In
Fig.~\ref{fig:echelle_diag}, we present a comparison of the frequencies
detected in this work with those detected in previous studies. This is
generally satisfactory agreement.  Some of our frequencies
are multiple, as indicated in Table~\ref{tab:freq}, which
could be attributed to the stellar rotation or to the modes being
resolved.

\subsection{Large and small separations}

\begin{figure*}
   \resizebox{\hsize}{!}{\includegraphics{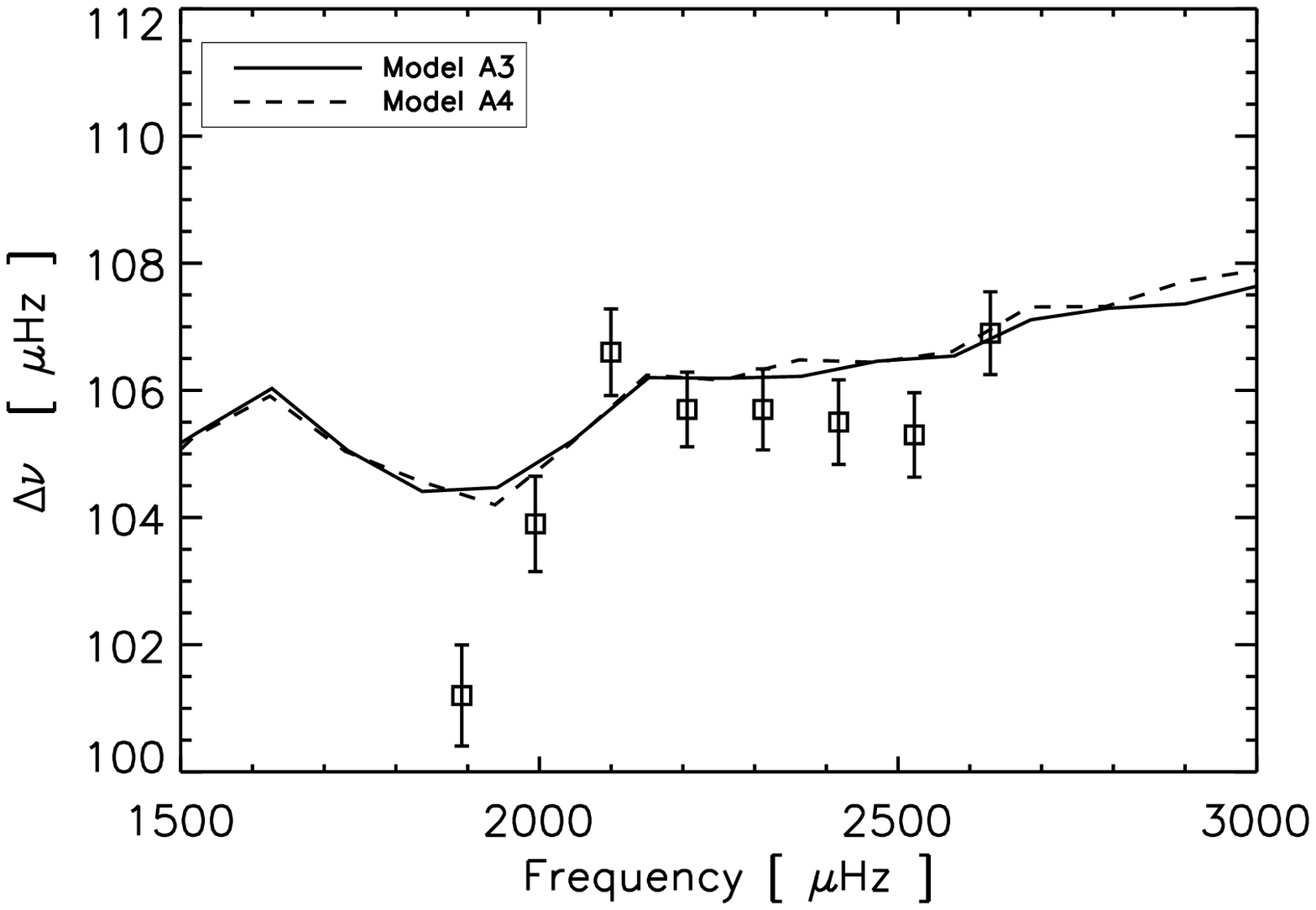} 
   \includegraphics{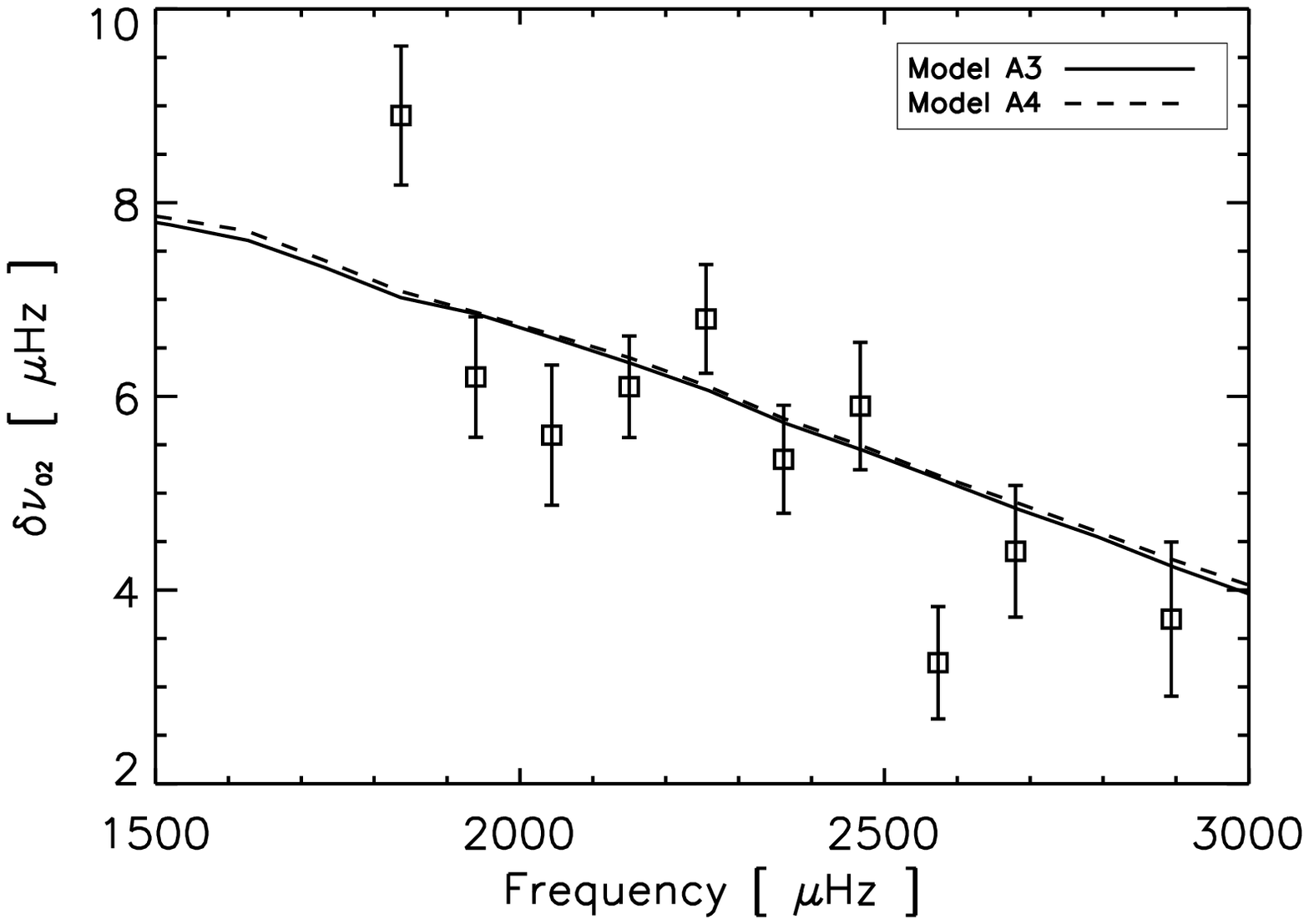}}
   \caption{Left panel: comparison of the large separation for $\ell=0$
   modes with the one derived by \citet{Miglio05} in their models $A3$,
   with radiative core and $A4$, with convective core. Right panel:
   comparison of the small separation with the models $A3$ and $A4$.  The
   agreement between the data points of our study and the models is
   significantly better than for the small separation obtained by
   \citet{Bouchy02} (see \citealt{Miglio05})}
   \label{fig:ls_ss_theorie}
   \end{figure*}

The left panel of Fig. \ref{fig:ls_ss_theorie} shows the large separation
as a function of frequency for the modes of degree $\ell = 0$.  To
compute the average large separation, $\Delta\nu_{0}$, we fitted a
linear relation $\nu_{n,\ell} = n \left( \Delta\nu_{0} + \epsilon \right)$,
where $\epsilon$ is a constant and the slope gives the mean
large separation. We also did this for each value of $\ell$ and made a weighted
sum to obtain $105.9 \pm 0.3 \: {\rm \mu Hz}$.  This is in good agreement with 
previously published values of $106.1 \pm 0.4 \: {\rm \mu Hz}$
\citep{Bedding04} and $105.9 \pm 0.3 \: {\rm \mu Hz}$ \citep{Bazot07}.

For the small separation, $\delta\nu_{02}$, we computed a mean of the values
shown in the right panel of the figure, and found $5.8 \pm 0.1 {\rm \mu
Hz}$ (the error bar is underestimated because it does not take into account
the additional error due to splittings).  Our value is slightly lower than
previously published values of $7.1 \pm 0.6 \: {\rm \mu Hz}$
\citep{Bedding04} and $6.9 \pm 0.4 \: {\rm \mu Hz}$ \citep{Bazot07}.  Note,
however, that our study spans a different and wider region in frequency
than that by \cite{Bazot07}, and our combined time series has higher
resolution than the UVES/UCLES data alone.  Our result is, however, in good
agreement with \citet{Bouchy02}.

\begin{table*}
\centering
\caption{\small Table of the amplitudes fitted by least squares fit (in
cm\,s$^{-1}$) and their S/N ratios associated. One can see that some modes
present a S/N ratio smaller than 3, which is the threshold of detection. In
fact the amplitude found for these modes by the pre-whitening algorithm was
higher but have decreased during the final least squares fit (global fit of
all detected frequencies together).} 
\begin{tabular}{ccccc|ccccccccccccccc}
\hline
\hline
 & & Amplitudes & & & & & S/N ratios & & \\
\hline
 & $\ell = 0$ & $\ell = 1$ & $\ell = 2$ & $\ell = 3$ & & $\ell = 0$ & $\ell = 1$ & $\ell = 2$ & $\ell = 3$ \\
\hline
n = 15 &          & $7.8$           & $11.1$/$12.6$          &               & n = 15 &		   & $3.0$     & $4.3/4.9$ & \\

n = 16 & $11.4$   & $15.7$          & $15.6$/$12.8$          & $6.8$         & n = 16 & $4.2$  & $6.1$     & $6.1/5.0$ & $2.6$ \\

n = 17 & $11.3$   & $11.3$          & $19.5$                 & $11.7$/$9.5$  & n = 17 & $4.4$  & $4.4$     & $7.6$     & $4.6/3.7$ \\

n = 18 & $9.5$    & $19.5$          & $15.6$/$12.4$          & $12.4$/$15.7$ & n = 18 & $3.8$  & $7.8$     & $6.1/4.8$ & $4.8/6.1$\\

n = 19 & $32.4$   & $21.1$          & $13.6$/$19.2$          & $11.8$/$12.0$ & n = 19 & $14.6$ & $8.2$     & $5.3/7.5$ & $4.6/4.7$ \\

n = 20 & $25.8$   & $31.0$          & $31.0$/$17.8$          & $7.2$/$9.2$   & n = 20 & $10.1$ & $12.1$    & $12.1/6.9$& $2.8/3.6$ \\

n = 21 & $19.9$   & $17.8$/$9.3$    & $21.8$                 & $15.6$        & n = 21 & $8.5$  & $6.9/3.6$ & $8.5$     & $6.1$ \\

n = 22 & $20.7$   & $12.4$/$18.7$   & $16.9$/$19.0$          & $13.4$        & n = 22 & $8.1$  & $4.8/7.3$ & $6.6/7.4$ & $5.2$ \\

n = 23 & $10.5$   & $26.3$          & $11.9$                 & $10.0$/$10.0$ & n = 23 & $4.1$  & $10.2$    & $4.6$	   & $3.9/3.9$ \\

n = 24 & $19.0$   & $9.3$/$22.8$    & $15.0$                 &               & n = 24 & $7.4$  & $3.6/8.9$ & $5.8$	   & \\

n = 25 &          & $11.4$          & $9.2$                  &               & n = 25 & 	   & $4.4$     & $3.6$ 	   & \\

n = 26 & $13.3$   & $9.3$           & $10.6$                 &               & n = 26 & $5.2$  & $3.6$     & $4.1$	   & \\

n = 27 &          & $8.2$           &                        &               & n = 27 & 	   & $3.2$     & 		   & \\

n = 28 &          & $10.1$          &                        &               & n = 28 & 	   & $3.9$     & 		   & \\

 \hline
 \end{tabular}
\label{tab:amplitude}
\end{table*}


\subsection{Rotational splittings and frequency}


\citet{Fletcher06} reanalysed the WIRE data obtained by
\citet{Schou00}. Fitting to the autocovariance function (ACF), they
obtained a value of $0.54 \pm 0.22 \: {\rm \mu Hz}$ for the rotational
splitting. They also performed a fit to the power spectrum and obtained
$0.64 \pm 0.25 \: {\rm \mu Hz}$, but suggested that this value is less
robust.  \citet{Bazot07} (based on the data of \citet{Pourbaix02} and
\citet{Saar97}) found $\Omega = 0.51 \pm 0.13 \: {\rm \mu Hz}$. However
from the $5$ splittings of $\ell = 2$ modes observed by \citet{Bazot07},
one can derive a rotational frequency of $0.75 \pm 0.22 \: {\rm \mu Hz}$,
significantly higher than the other values.

The visibilities of the multiplet components depend on the inclination
angle $i$ of the rotation axis of the star
(e.g.\citet{Gizon03,bal06,bal08}). We recall that a high value of the
inclination angle leads to the peaks with $m = \pm \ell$ being higher than the
others in the multiplet.
%
%
\citet{Bouchy02} adopted an inclination of $i = 79$ $^{\circ}$
\citep{Pourbaix02} to estimate the visibilities of the split modes.  They
assumed that the inclination of the star and of the orbit are the same,
which is not necessarily the case.  Their results for the visibilities are
given in Table~\ref{tab:ampli_splitting}, from which we see that
the two most visible peaks of a split $\ell =1$ mode are those with $m =
\pm 1$.  Thus, if we take a value of $0.5$ ${\rm \mu Hz}$ for the rotation
frequency, these peaks are expected to be separated by $1$ ${\rm \mu Hz}$,
and will have equal amplitudes. For an $\ell =2$ mode, the distance
between the two major peaks is $2$ ${\rm \mu Hz}$ and the quantum numbers
are $m = \pm 2$, while for an $\ell = 3$ mode, the distance is $3$ ${\rm \mu
Hz}$ and $m = \pm 3$. That is what we would expect to find for such a
rotation frequency.

\begin{figure}
   \resizebox{\hsize}{!}{\includegraphics{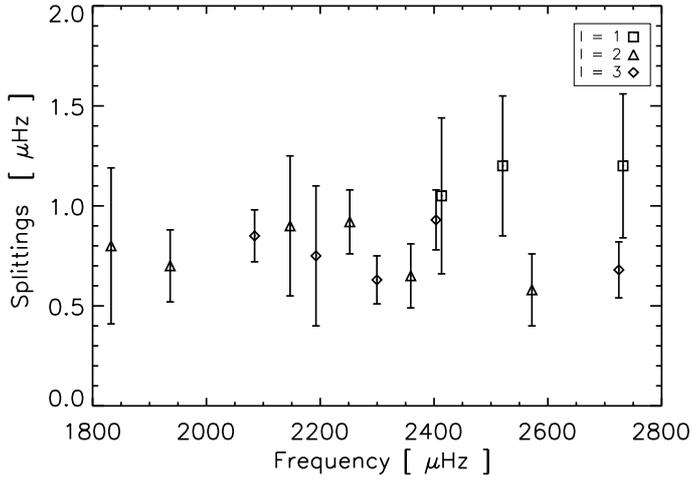}}
   \caption{Size of the splittings for $\ell = 1,2$ and $3$ modes, divided
   by $2m$. The mean of these splittings should give the value of the
   rotation frequency of the star (see text).}
   \label{fig:splitting}
   \end{figure}

We have detected 14 possible splittings, listed in Table
\ref{tab:freq}.  They are shown in Fig. \ref{fig:splitting},
normalized according to their most probable $m$
quantum number. However, note that one should be cautious with the $\ell =
1$ splittings. Firstly, they could be caused by the finite mode lifetime ($3.9 \pm 1.4$
days according to \citet{Fletcher06} and $2.3_{-0.4}^{+1.0}$ days
according to \citet{Kjeldsen05}). Secondly, amplitude ratios between prograde ($m = +1$)
and retrograde ($m = -1$) modes are found to be approximately a factor two,
which is not consistent with expected mode visibility ratios.
%
%
For the modes of higher angular degree, the frequency separation between $m
= +\ell$ and $m = -\ell$ mode is higher than the mode line-width inferred
by both \citet{Kjeldsen05} and \cite{Fletcher06}. In addition, mode
amplitude ratios are compatible with mode visibility ratios (see table
\ref{tab:amplitude} and \ref{tab:ampli_splitting}), except for the
$2358.3$/$2359.6$ mode, which is therefore not a good candidate of
splitting.


We computed a rotational frequency of $0.77 \pm 0.05 \: {\rm \mu Hz}$ by
taking the weighted mean of all splittings. If we exclude the $\ell =
1$ splittings, which have values close to the
limit of resolution, a value of $0.75 \pm 0.05 \: {\rm \mu
Hz}$ is found.  This last value will be kept as more secure for the reasons
mentioned above.
The rotation frequency we have found, although greater, is within one sigma
of both values derived by \citet{Fletcher06}. The value given by
\citet{Bazot07}, $\Omega = 0.51 \pm 0.13 \: {\rm \mu Hz}$, is only based on
spectroscopic data \citep{Pourbaix02, Saar97}. From their identified $\ell
= 2$, we can derive a splitting of $0.75 \pm 0.22 \: {\rm \mu Hz}$, in
good agreement with our value. One has to keep in mind that our value,
$0.75 \pm 0.05 \: {\rm \mu Hz}$, is based on measurements of the rotational
frequency by means of an identification of split modes. 
Note that this value indicates a
significantly faster rotation rate than the one determined by \citet{jay}
and \citet{Saar97}, who found a period between 20 and 30 days. Our
rotational frequency, if correct, corresponds to a period of about 15\,days.

\begin{table}
\centering
\caption{\small Table of the expected amplitudes ratios of split modes, in
function of $\ell$ and m quantum numbers, compared to the amplitude of the
non-split mode ($\ell=0$,$m=0$)}.  
\begin{tabular}{ccccccccccccc}
\hline
\hline
 & $\ell,m$ & amplitudes ratios\\
\hline
& $\ell$ = 0, m = 0    & 1.00\\
& $\ell$ = 1, m = 0    & 0.25\\
& $\ell$ = 1, m = -1,1 & 0.90\\
& $\ell$ = 2, m = 0    & 0.40\\
& $\ell$ = 2, m = -1,1 & 0.21\\
& $\ell$ = 2, m = -2,2 & 0.53\\
& $\ell$ = 3, m = 0    & 0.09\\
& $\ell$ = 3, m = -1,1 & 0.12\\
& $\ell$ = 3, m = -2,2 & 0.08\\
& $\ell$ = 3, m = -3,3 & 0.18\\
 \hline
 \end{tabular}
\label{tab:ampli_splitting}
\end{table}

%

\section{Comparison with models}


A great number of theoretical studies dealing with $\alpha$ Cen A have been
published in the recent years, in particular since solar-like oscillations
were first detected by \citet{Bouchy02} \citep[see e.g.][]{Thevenin02, Thoul03,
Eggenberger04, Miglio05, Yildiz07}. The A component of the binary system is
indeed particularly interesting to model because its mass, $1.105 \pm 0.007
\: {\rm M_{\odot}}$ \citep{Pourbaix02}, is close to the limit above which
stars in the main sequence phase (MS) keep the convective core they
developed on the pre-MS (PMS).

\begin{figure}
   \resizebox{\hsize}{!}{\includegraphics{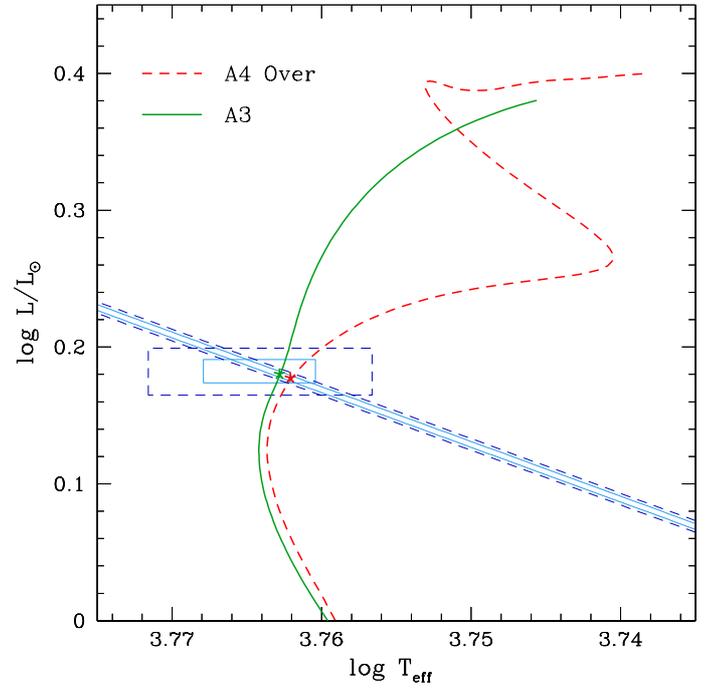}}
   \caption{HR diagram showing the evolutionary tracks of model A3 (with a
radiative core) and model A4 (with a convective core) from
\citet{Miglio05}. The error boxes for $T_{\rm eff}$, $\log L/L_{\odot}$,
and radii correspond to 1$\sigma$ (solid line) and $2\sigma$
(dashed-line).}
   \label{fig:HR_track}
   \end{figure}

\citet{Miglio05} performed several calibrations of the $\alpha$ Centauri
system to fit both classical constraints (photometric, spectroscopic and
astrometric) as well as the asteroseismic constraints given by
\citet{Bouchy02} and \citet{Carrier03}.  They found that
models with either a radiative or a convective core could reproduce equally well
classical constraints, as we show in Fig. \ref{fig:HR_track}.  An
overshooting parameter $\alpha_{\rm OV}>0.15$\footnote{The extension of the
overshooting layer in the models is defined as $ov=\alpha_{\rm OV}\,{\rm
min}(H_{\rm p}(r_{\rm c}),r_{\rm c})$, where $H_{\rm p}$ is the pressure
scale height and $r_{\rm c}$ the radius of the convective core} was found
to be sufficient in the calibrations for a convective core to persist after
the PMS.  We see in Fig. \ref{fig:HR_track} that the effect of
overshooting is clearly important in the evolution of the star: the
evolutionary track of model A4 (computed with $\alpha_{\rm OV}=0.2$) has a
similar behaviour to that of more massive stars, where a convective core is
present regardless of the amount of overshooting adopted.

Despite their similar photospheric constraints, models A3 and A4 have
significantly different interiors. While model A3 has a
radiative energy-generating core characterized by a smooth chemical
composition gradient, model A4 presents an adiabatically stratified
fully-mixed region extending to about $10\%$ the total mass of the star,
and a sharp chemical composition gradient at its edge. This different
structure, as shown by \citet{Miglio05}, leaves a clear signature in the
oscillation frequencies, in particular in the asteroseismic quantities $r_{10}$
and $r_{02}$ \citep{Roxburgh03}:
\begin{equation}
r_{02} = \frac{\delta \nu_{02}}{\Delta \nu_1}   \ ; \ \ \
r_{10} = \frac{\delta \nu_{10}}{\Delta \nu_0}  \ ,
\end{equation}
with the small separation between $\ell = 1$ and $\ell = 0$ modes defined
as:
\begin{equation}
\delta \nu_{10} = \frac{1}{8}\left(\nu_{n-1,1} - 4 \nu_{n,0} +6 \nu_{n,1}
-4 \nu_{n+1,0} + \nu_{n+1,1}\right) \ .
\end{equation}
Note we have adopted the 5-point definition for $\delta \nu_{10}$ proposed
by \citet{Roxburgh03}, which is smoother than the conventional 3-point
separation.  These ratios are known to be largely independent of outer
layers of the star, and provide a reliable probe of the near-core
structure.
However, the uncertainties on the oscillation frequencies prevented 
\citet{Miglio05} from drawing firm conclusions on the core properties of $\alpha$
 Cen A. The sensitivity of $\delta \nu_{10}$ to the properties of the core
 has also been discussed by \citet{deheuvels}.


   \begin{figure*}
   \resizebox{\hsize}{!}{\includegraphics{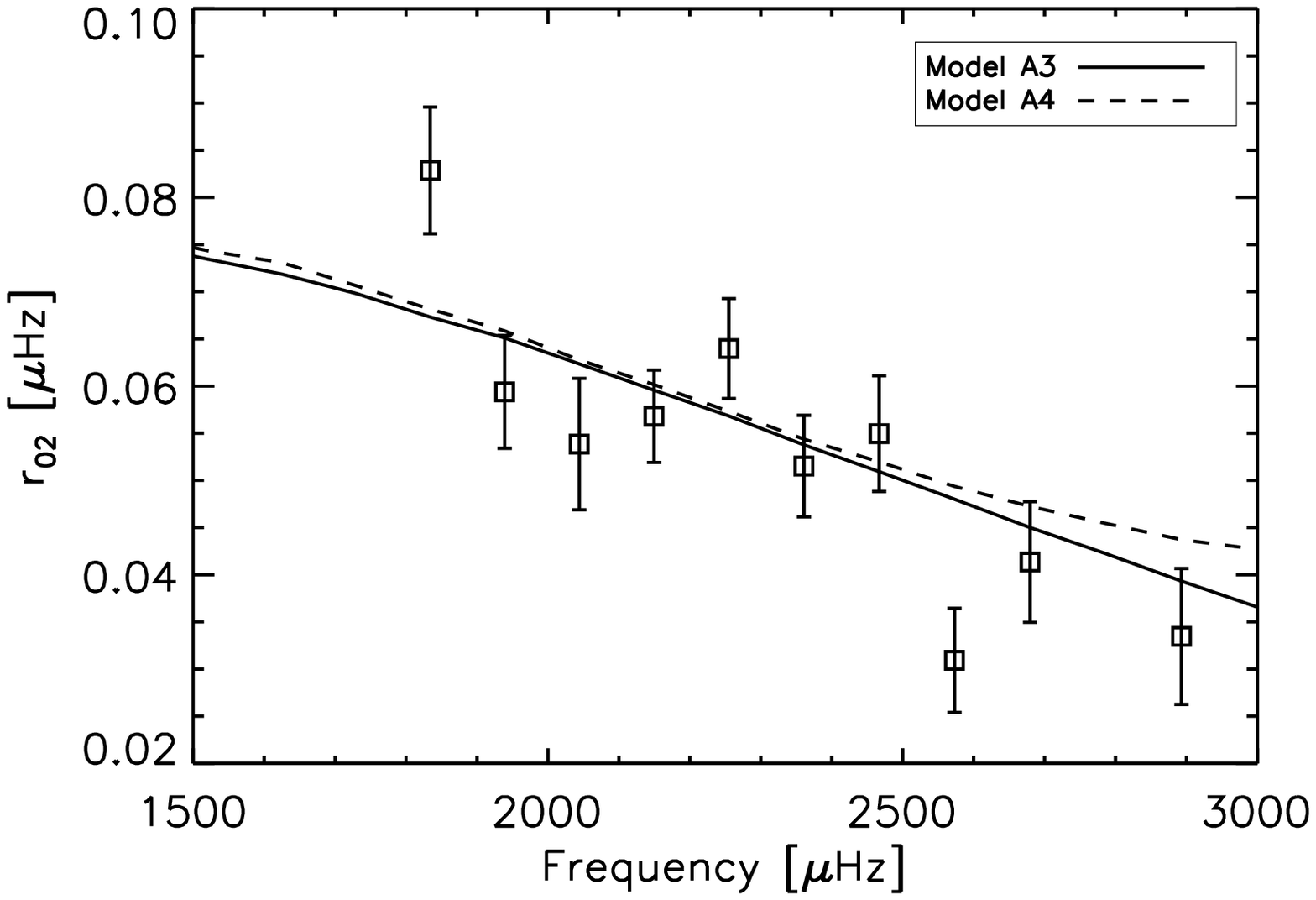}
   \includegraphics{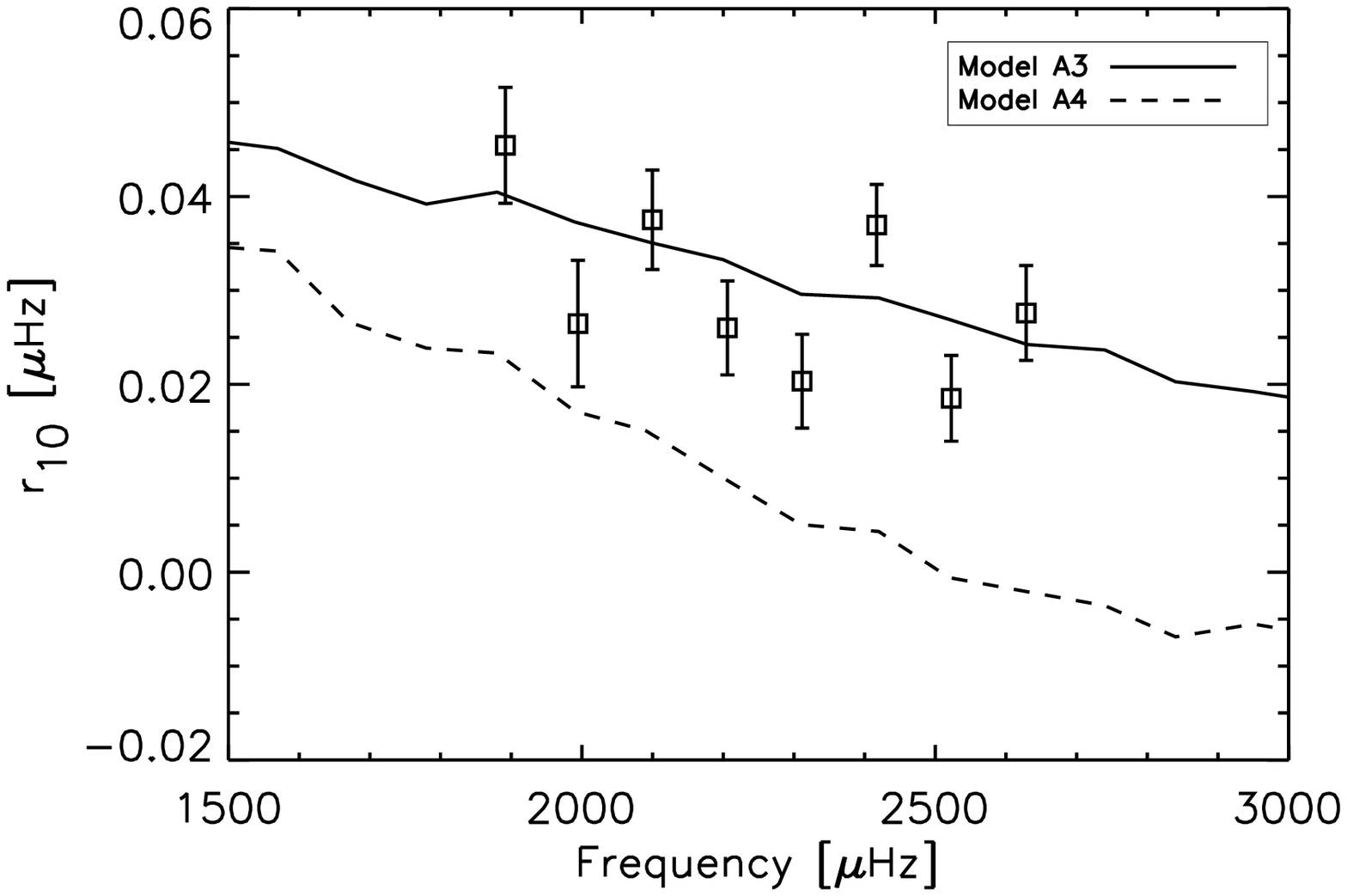}}
   \caption{Left: comparison of the $r_{02}$ constraint with the models
   $A3$ and $A4$. Right: comparison of the $r_{10}$ constraint with the
   models $A3$ and $A4$. The data points seem clearly to agree with the
   model A3, a fact that was not established with the data of
   \citet{Bouchy02}. Indeed, all their points for the $r_{10}$ asteroseismic
   constraint were situated between the curves of the models A3 and A4 (see
   \citet{Miglio05}).}
   \label{fig:r02_r10_theorie}
   \end{figure*}
   \begin{figure}
   \resizebox{\hsize}{!}{\includegraphics{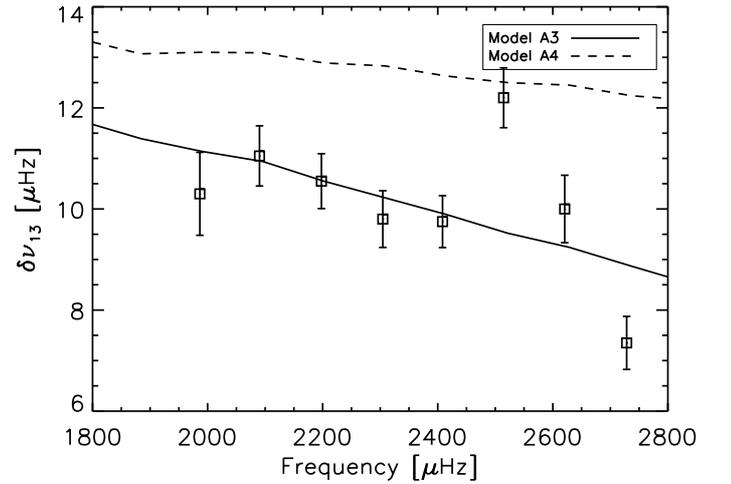}}
   \caption{Comparison of the $ \delta\nu_{13}$ constraint with the models
   $A3$ and $A4$. The model without overshooting is in a better agreement
   with the data than the model with overshooting.}
   \label{fig:delta_nu_13_theorie}
   \end{figure}

In Figures \ref{fig:r02_r10_theorie} and
\ref{fig:delta_nu_13_theorie} we compare the frequency separations $r_{10}$, $r_{02}$
and $\delta\nu_{13}$ derived by our study with those of models A3 and
A4. Note that when two peaks were identified for a given mode $\nu_{n,\ell}$,
we assumed that they are modes with opposite $m$-degrees and took the
average to get the mode $\nu_{n,\ell,0}$. When only one peak was detected
for a given non-radial mode, it was assumed to be $m$\,=\,0, which
leads to possible bias that may explain some of the outliers of
Figs.~\ref{fig:ls_ss_theorie},~\ref{fig:r02_r10_theorie}
and~\ref{fig:delta_nu_13_theorie}. Although $r_{02}$ 
does not discriminate between models with radiative (A3) and with
convective (A4) core, $r_{10}$ clearly favours model A3, a fact that could
not be established with the data of \cite{Bouchy02} in the work of
\cite{Miglio05}. Moreover, the availability of $\ell = 3$ modes in the
power spectrum also allows such a discrimination via the
small separation
$\delta\nu_{13}$: this comparison further supports the conclusion that a
model with a convective core having about 10\% of the stellar
mass, such as model A4, can be rejected on the basis of the asteroseismic
constraints.



\section{Summary}

Observations of $\alpha$ Centauri A allowed us to derive an accurate set of
asteroseismic constraints to compare with models and make inferences on the
internal structure of our closest stellar neighbour.  We combined the time
series obtained by three spectrographs during a multi-site observation
campaign carried out in 2001. While the combined time series is as long as
the one of \citet{Bouchy02}, it contains almost $5$ times more spectra, and
daily aliases are reduced by a factor $2.6$.   These
improvements in the time series allowed us to detect 44 frequencies with
$\ell = 0,1,2,3$ by means of a pre-whitening algorithm, of which 14 showed
possible rotational splittings. Five of these splittings have been obtained
for modes $\ell = 3$, the first time in $\alpha$ Cen A. A new \'echelle
diagram has been derived and is in overall good agreement with the results
of the previous analyses. New values of the large and small separations
have been derived from the set of frequencies.

A comparison with the stellar models by \citet{Miglio05} indicates that the
asteroseismic constraints determined in this study (namely $r_{10}$ and
$\delta\nu_{13}$) allow us to set an upper limit on the amount of
convective-core overshooting needed to model stars of mass and metallicity
similar to those of $\alpha$ Cen A. As described in section 4, a model of
the star with a radiative core represents the observed $r_{10}$ and
$\delta\nu_{13}$ separations significantly better than the model with
a convective core.\\


\begin{acknowledgements}
FC is a postdoctoral fellow of the Fund for Scientific Research, Flanders
(FWO).  AM is a postdoctoral researcher of the 'Fonds de la recherche
scientifique' FNRS, Belgium.  
This work was also supported financially by the Australian Research Council
and the Danish Natural Science Research Council.

\end{acknowledgements}

\bibliographystyle{aa}
\bibliography{memoire_biblio}

\end{document}